\documentclass[12pt,fleqn,psfig]{article}
\usepackage{amsmath, amssymb, graphicx, float}

\begin{document}

{\hfill \parbox{4cm}{
        Brown-HET-1563 \\
}} \vskip100pt
%%%%%%%%%%%%%%%%   TITLE    %%%%%%%%%%%%%%%%%%%%

\thispagestyle{empty}
\renewcommand{\thefootnote}{\fnsymbol{footnote}}

\begin{center} \noindent \Large
Remarks on Power Spectra of Chaotic Dynamical Systems
\end{center}

\bigskip\bigskip\bigskip

\bigskip\bigskip\bigskip

\centerline{G.~Guralnik\footnote[1]{gerry@het.brown.edu},
Z.~Guralnik\footnote[2]{zack@het.brown.edu},
C.~Pehlevan\footnote[3]{cengiz@het.brown.edu}}

\bigskip

\centerline{Department of Physics} \centerline{Brown University}
\centerline{Providence, RI 02912}
%\bigskip
%\centerline{$^b$ ??}
%\bigskip\bigskip

\bigskip\bigskip\bigskip\bigskip

\renewcommand{\thefootnote}{\arabic{footnote}}

\centerline{ \small Abstract}
\bigskip

\small We develop novel methods to compute auto-correlation
functions, or power spectral densities,  for chaotic dynamical
systems generated by an inverse method whose starting point is an
invariant distribution and a two-form. In general, the inverse method makes some aspects of chaotic dynamics calculable by methods familiar in quantum field theory. This approach has the numerical advantage of being amenable to Monte-Carlo parallel computation. We demonstrate the approach on a specific example, and show how auto-correlation functions can be computed without any direct numerical simulation, by Pade approximants of a short time expansion. \vskip40pt
\newpage

\section{Introduction}

Due to their sensitive dependence on initial conditions,  the
precise behavior of a chaotic dynamical system over long times is
unpredictable in practice, even though the underlying system is
deterministic. However there is no fundamental obstruction to
calculating statistical properties of the long time behavior of
chaotic systems.

Typically, statistical properties of chaotic systems are calculated
by direct numerical simulation,  with the assumption that the
statistics, in the form of time averages, converge quickly compared
to the rate at which numerical errors accumulate due to the
sensitivity to initial conditions.  Direct numerical simulation of
systems with a very large number of degrees of freedom requires
significant computational power which in many cases is not yet
available.  Moreover,  it is difficult to gain theoretical insight
from direct simulations.

In a previous paper \cite{Zach}, which we review in section II, we
described an alternative approach to direct numerical simulation.
This approach involves an inverse method whereby chaotic dynamical
systems are generated, given a two-form and exact statistical
information given by an invariant distribution over phase space.
Several examples of chaotic systems were given for which there was
precise agreement between equal time moments calculated by direct
numerical simulation or by using the initial invariant distribution.

There are several advantages to the inverse approach over direct
simulation. In principle, the inverse approach can be applied to
systems with a large number of degrees of freedom, without placing
intense demands on computational power.  It is also possible to
begin a classification of chaotic systems in terms of analytic
properties of the statistics.  We expect that, using the inverse
approach, many theoretical insights into properties of chaotic
dynamical systems are forthcoming.

In \cite{Zach},  only equal time moments (and cumulants) were
computed for chaotic systems generated by the inverse approach. The
present work is a sequel to \cite{Zach}, in which we will apply the
inverse method to the computation of auto-correlation functions,
whose Fourier transform is the power spectral density. Ordinarily,
the auto-correlation or power spectrum is computed by a direct
numerical simulation with a long run-time. However, for chaotic
systems generated by the inverse method, the auto-correlation can be
computed by a Monte-Carlo simulation, involving many fixed time
simulations from different initial conditions, as opposed to one
long simulation. This has the great advantage of being amenable to
parallel computation.  We discuss the Monte-Carlo approach in
section III.  Although we only consider auto-correlation functions
here, we emphasize that our approach is applicable to any
correlation functions of phase space variables at different times.

For chaotic systems generated by the inverse method, it is also
possible to calculate auto-correlation functions without direct
numerical simulation.  In section IV, we show that the inverse
approach yields a small $t$ expansion for the auto-correlation
$G(t)$, whose validity can be extended deeper into the complex $t$
plane by Pade approximants. We shall see that this approach shows
good agreement with results obtained by direct numerical simulation.

Power spectra of dynamical systems are sometimes obtained from
Fourier decomposition of a signal within some time window. For a
chaotic system, this approach is rife with ambiguities and pitfalls,
unlike the Fourier transform of the auto-correlation function which
is well defined.  However these pitfalls are themselves interesting
and can be well understood using the inverse approach, as we shall
see in section V.

%Of course, the inverse method presents several difficulties which
%must be surmounted. For instance, while is easy to generate many (in
%fact an infinite number) of chaotic systems for which statistical
%information is known exactly, it is generally non-trivial to reverse
%engineer a specific system of physical interest.

\section{Review of the inverse approach}

We give a very brief review of the inverse method here.  For details
the reader is referred to \cite{Zach}.  We consider deterministic
dynamical systems, with phase space coordinates $x_1(t)\cdots
x_N(t)$ and equations of motion
\begin{align}\label{fp}
\frac{dx_n}{dt} = v_n(x_1,\cdots x_N)
\end{align}
A distribution over initial conditions $\rho[x]$ which is left
invariant by the time evolution is a solution of the zero diffusion
limit of the static Fokker-Planck equation,
\begin{align}
\vec\nabla\cdot (\rho \vec v) =0\, ,
\end{align}
For $N=3$, this implies that $\rho\vec v$ can be written as the curl
of a vector field,
\begin{align}\vec v= \frac{\vec\nabla\times\vec A}{\rho}\, ,
\end{align}
and the inverse method amounts to choosing $\vec A$ and $\rho$, such
that $\vec v$ is polynomial in the phase space coordinates $x_n$.
For $N>3$, it is convenient to use the language of differential
forms, in which case \eqref{fp} becomes
\begin{align}
d{}^*(\rho v) =0
\end{align}
where $v$ is the velocity one form $v_m dx^m$, $d$ is the exterior
derivative, and * indicates the Hodge dual.  This implies that
\begin{align}
v=\frac{{}^*d{\cal A}}{\rho}
\end{align}
where ${\cal A}$ is an $N-2$ form.  The problem is then to choose
$\rho$ and ${\cal A}$ such that $v$ is polynomial in $x_n$.  This
was done in \cite{Zach} for a few increasingly complex analytic
structures of $\rho$ and ${\cal A}$.

Since ${}^*{\cal A}$ is a two-form,  which is more conveniently
written down for large $N$ than ${\cal A}$,  it can be said that an
invariant distribution and a two-form uniquely specify dynamical
systems.  This is very similar to the specification of Hamiltonian
dynamical systems by a Hamiltonian and a symplectic form, although
far more general.  For example $N$ need not be even, and a Liouville
theorem, $d{}^*v=0$, need not apply.  There also are not necessarily
any conserved quantities.

Amongst the simplest distributions, so far as analytic structure is
concerned,  are polynomial distributions for which the real zeroes
form a closed manifold inside of which the polynomial is positive.
For these,
\begin{align}
{\cal A} = \rho^2 \Omega
\end{align}
were $\rho$ is a polynomial, and $\Omega$ is a polynomial $N-2$
form, so that
\begin{align}\label{rep}
v={}^*(\rho d\Omega+ 2 d\rho\wedge\Omega)
\end{align}
Chaotic dynamical systems of this type are invariably repellers (see
\cite{Zach}). Chaotic attractors arise from distributions and
two-forms with more complicated analytic structure, which are also
discussed in \cite{Zach}.

Although  a distribution $\rho$ satisfying \eqref{fp} is an
invariant distribution over initial conditions,  it does not
necessarily describe the statistics of a single chaotic trajectory.
It must be verified that the dynamics is chaotic and that $\rho$, or
strictly speaking its projection, is an ergodic measure, having no
convex decomposition into independent invariant measures $\rho \ne
x\rho_1 + (1-x)\rho_2$. In many instances it is necessary to modify
the domain of support of the initial distribution $\rho$ to make it
ergodic.  For chaotic invariant sets with fractional information
dimension,  a function $\rho(x)$, with support in $N$ dimensions,
can not be ergodic, although it may still contain exact information
about the statistics, either via projection to lower integer
dimension or, after Fourier transformation,  as the generator of
polynomial moments.  The question of the general relevance of the
distributions $\rho$ arising in the inverse approach has not been
definitively resolved.  However, for the chaotic systems which have
been considered so far in \cite{Zach},  the distributions used to
generate the dynamical systems give extremely accurate results for
moments and cumulants (connected equal time correlation functions).
Of course it is possible that these systems have information
dimension which is very nearly $N$,  in which case $\rho$ yields, at
worst, an extremely accurate approximation to the exact statistics.

We emphasize that the results in this article are not restricted to
chaotic dynamical systems generated by the inverse method.  In fact,
they may be applied to any system for which there is known
information about the invariant measure over phase space or equal
time moments.  The point of this article is to use known information
about static (time independent) statistics to facilitate the
computation of time dependent quantities, such as auto-correlation
functions and power spectral densities.

\section{Monte-Carlo computation of the auto-correlation and power
spectrum}

Power spectra are useful signatures of dynamical systems derived
from time series data.  The power spectrum associated with a signal
$x(t)$ is defined as the squared amplitude of the Fourier transform
of $x(t)$. However, the Fourier transform of an a-periodic chaotic
signal $x(t)$ is not generally well defined since $x(t)$ does not
fall of as $t\rightarrow\pm\infty$.  The power spectral density on
the other hand is well defined;
\begin{align}\label{pspec}
\tilde G(\omega)\equiv \int_{-\infty}^{\infty}\, d\Delta\,
\exp(i\omega \Delta) G(\Delta)
\end{align}
where $G(\Delta)$ is the auto-correlation function,
\begin{align}
G(\Delta) \equiv \lim_{T\rightarrow\infty}
\frac{1}{T}\int_{-T/2}^{T/2}\,dt\, x(t)x(t+\Delta)\, .
\end{align}
Due to the tendency of chaotic systems to `forget' their initial
conditions,  $G(\Delta)$ falls off with large $\Delta$ (assuming
$<x>=0$) sufficiently rapidly that the Fourier transform
\eqref{pspec} is well defined.  The total power between frequencies
$\omega$ and $\omega+d\omega$ is given by $\tilde G(\omega)d\omega$.

A remarkable property of chaotic systems is that time averages of
functions of phases space over a trajectory are equal to spatial
averages with respect to an ergodic measure: \begin{align}
<f>=\lim_{T\rightarrow\infty}\frac{1}{T}\int_0^T\,dt\, f(\vec x(t))
= \int d\mu(\vec X) f(\vec X)\end{align} The auto-correlation of a
phase space variable $x_i$ can therefore be written as
\begin{align}\label{theqn}
G_i(\Delta)=\int d\mu(\vec X) f_\Delta(\vec X)
\end{align}
where \begin{align}\label{fps} f_\Delta(\vec X) \equiv
x_i(t=0)x_i(t=\Delta)\, , \,\,\,\,\, {\rm with}\,\, x_i(0)= X_i\,
.\end{align}  Thus, when the exact invariant measure is known, the
auto-correlation and power spectral density can be computed by
Monte-Carlo methods, which entail multiple fixed time simulations of
the dynamical system from randomly generated initial conditions
rather than a single very long simulation.  This approach is
amenable to efficient parallel computation schemes.

\section{An example}

Consider a specific chaotic system of the type \eqref{rep} (studied
in \cite{Zach}) with
\begin{align}\label{deft} \rho &=1-x^4-y^2-z^6 \nonumber \\
\Omega &= (xyz)dy + (y^2) dz\, .
\end{align}
yielding
\begin{align}\label{dyns}
&v_x=13z^6xy-6y^3-xy+2y+yx^5-2yx^4-2yz^6+y^3x\, , \nonumber \\
&v_y=8x^3y^2\, , \\
&v_z=-9x^4yz+yz-yz^7-y^3z \, . \nonumber \end{align} Initial
conditions with $\rho(x,y,z)>0$, $y>0$ and $z>0$ give rise to
chaotic trajectories whose statistics are described by an invariant
distribution $\tilde \rho = \rho$ for $\rho(x,y,z)>0,\, y>0,\, z>0$,
up to a normalization, and $\tilde \rho =0$ outside this domain.

Next we will evaluate the auto-correlation $G_x(\Delta)=
<x(t)x(t+\Delta)>$ for these chaotic trajectories.  The
auto-correlation can be evaluated by direct numerical simulation,
taking the time average of $x(t)x(t+\Delta)$ over a long run-time.
Alternatively,  it can be evaluated using \eqref{theqn},\eqref{fps};
\begin{align}\label{example}
G_x(\Delta) = \int dX dY dZ \tilde\, \tilde\rho(X,Y,Z)\,
f_\Delta(X,Y,Z)
\end{align}
where
\begin{align}
F_\Delta(X,Y,Z) = x(t=0)x(t=\Delta)\end{align} with \begin{align}
x(t=0)=X,\,\,y(t=0)=Y,\,\,z(t=0)=Z
\end{align}
Evaluating \eqref{example} by Monte Carlo simulation amounts to
generating initial conditions $X,Y,Z$ randomly according to the
distribution $\tilde\rho$, and simulating the time evolution from
each of these points over a duration $\Delta$. The results of both
the Monte-Carlo calculation and a direct long-duration numerical
simulation of \eqref{dyns} are shown in figure 1, showing extremely
good agreement.

The example we have given is a small dimension system, with $N=3$,
so there is little computational advantage to the Monte-Carlo
approach over a direct long duration simulation.  The advantage
comes at large $N$, for which the Monte-Carlo approach will be far
faster, as it is amenable to parallel computation.

%As discussed \cite{Zach}, dynamical systems can be uniquely
%characterized by an invariant distribution over phase space
%$\rho(\vec X)$ and a two-form $\Omega^{(2)}(\vec X)$. In many cases,
%$\rho(\vec X)$ is closely related to an ergodic measure,  and seems
%to give extremely accurate (if not exact) results for cumulants of
%chaotic trajectories. There are presently unresolved subtleties
%regarding $\rho(\vec X)d^N\vec X$ as the invariant measure of a
%chaotic set, since it is not consistent with a fractional integer
%dimension $<N$. Nevertheless, it has been suggested that this
%distribution may yield exact results for a subset of cumulants, or
%upon projection to lower integer dimension. If the fractional
%information dimension is very nearly $N$,  then $\rho(\vec X)$ may
%still at least provide a very good approximation to an ergodic
%measure. Given the success of the inverse approach of \cite{Zach},
%we will proceed to apply it here to compute auto-correlations and
%power spectral densities using \eqref{theqn}.

\begin{figure}[H]\centerline{
\includegraphics[width=50mm,height=50mm]{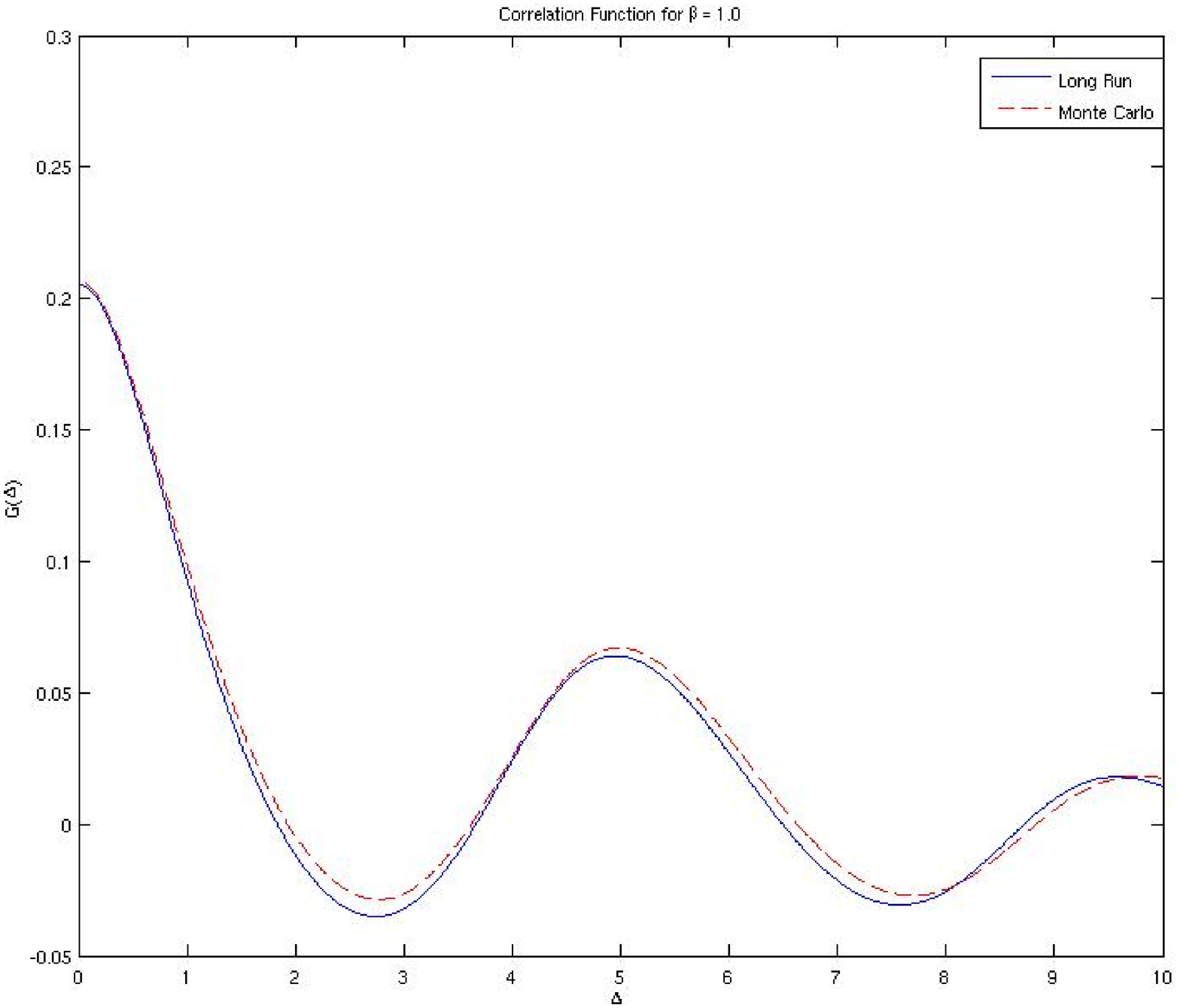}\qquad
\includegraphics[width=50mm,height=50mm]{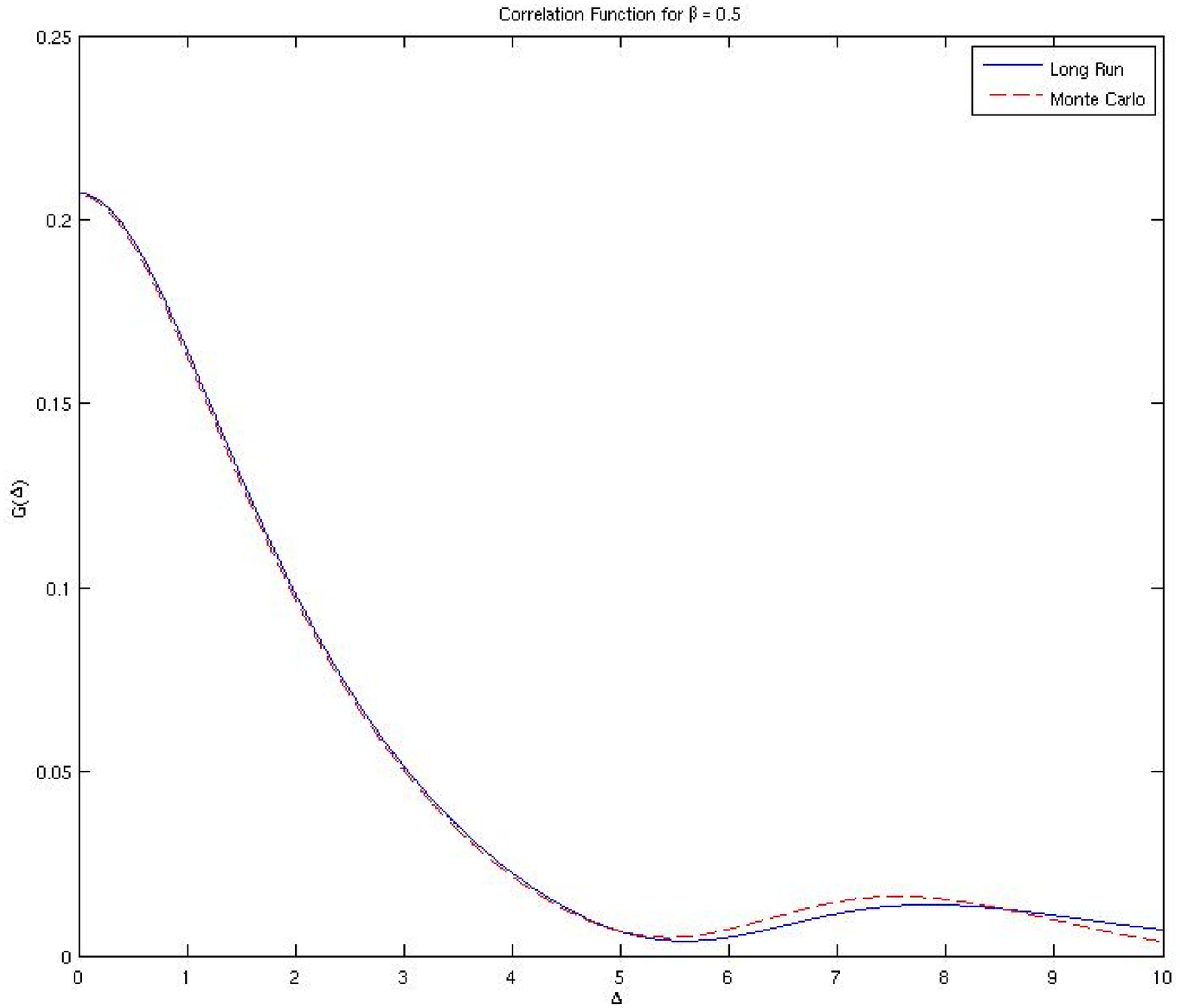}}\centerline{
\includegraphics[width=50mm,height=50mm]{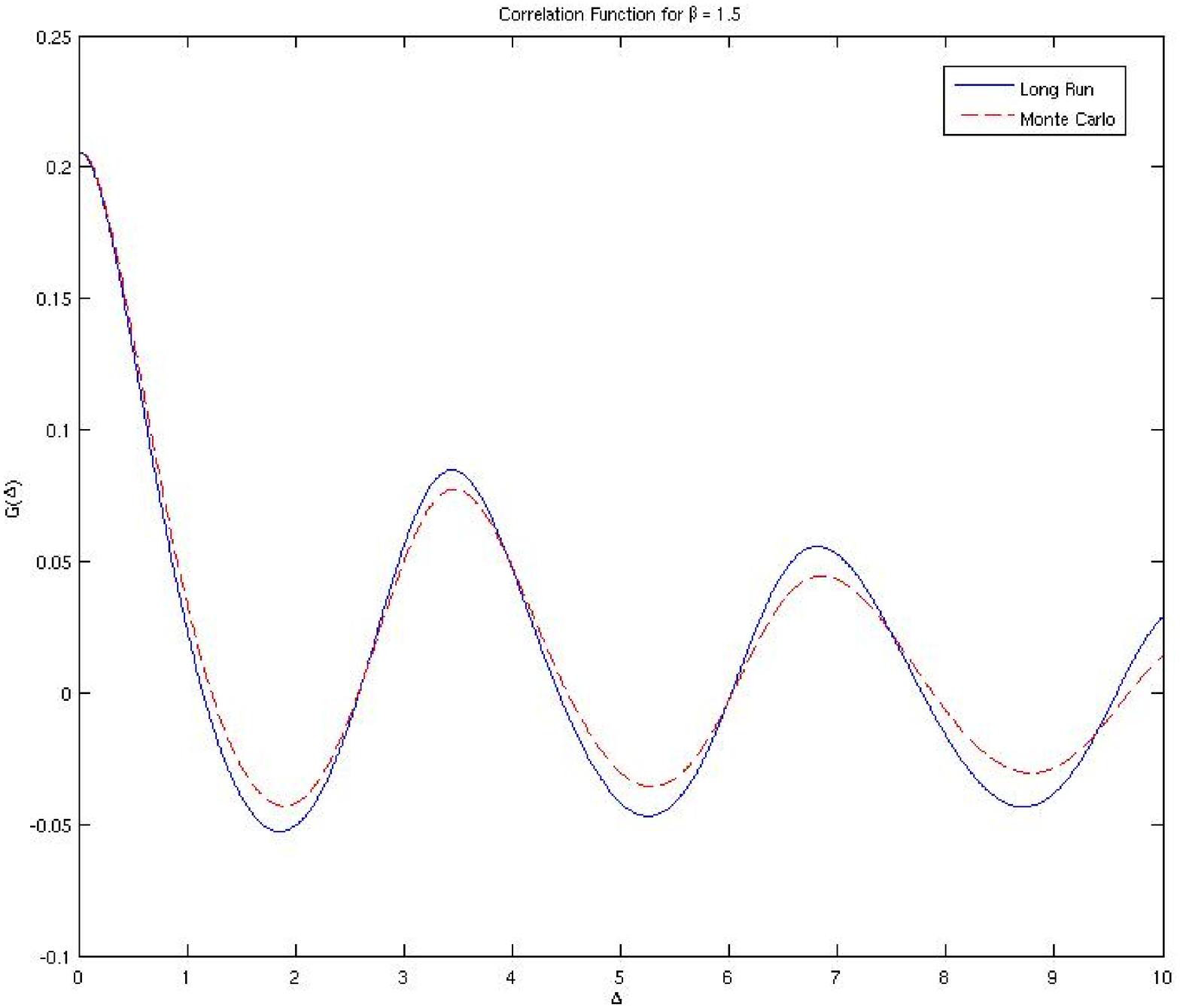}} \caption{Auto-correlation functions calculated by Monte-Carlo
simulation (dashed line) and direct long-duration simulation (solid
line). The dimensionless number $\beta$ parameterizes a variation of
the dynamical system \eqref{dyns}, obtained from $\rho
=1-x^4-y^2-z^6,\,\,\, \Omega = (xyz)dy + \beta(y^2) dz$. }
\end{figure}

\section{Calculating the auto-correlation without direct\\ numerical simulation.}

Using \eqref{theqn}, the auto-correlation function $G_i(\Delta)$ can
be expanded as a power series in $\Delta$.  To do so let us first
expand \eqref{fps} in $\Delta$:
\begin{align}
F_{\Delta}(\vec X) \sim x_i(0)\left(x_i(0)+\Delta \dot x_i(0)
+\frac{1}{2} \Delta^2 \ddot x_i(0)\cdots \right)\, , \,\,\,{\rm
with}\,\,\, \vec x(0)=\vec X\,.
\end{align}
Note that $x_i\frac{d^nx_i}{dt^n}$ can be written as a total time
derivative for odd $n$: \begin{align}x_i \frac{dx_i}{dt}  =
\frac{d}{dt}\left(\frac{1}{2}x_i^2\right),\,\,\,
x_i\frac{d^3x_i}{dt^3} =
\frac{d}{dt}\left(x_i\frac{d^2x_i}{dt^2}-\frac{1}{2}\left(\frac{dx_i}{dt}\right)^2\right)\,,\,\,{\rm
etc}\end{align} For even $n$, $n=2m$,
\begin{align} x_i\frac{d^nx_i}{dt^n} =
\frac{d}{dt}\left(g_n\right) +(-1)^{m}\left(\frac{d^m
x_i}{dt^m}\right)^2
\end{align}
where we will not bother to specify $g_n$, since total time
derivatives vanish when averaged over a chaotic
trajectory\footnote{We are assuming a bounded system with no
explicit time dependence in the equations of motion.}. Thus
\begin{align}\label{sers}
G(\Delta)=\left<F_\Delta\right> \sim \sum_{m=0}^\infty
\alpha_m\Delta^{2m}
\\ \alpha_m \equiv (-1)^m\frac{1}{2m!}\left<\left(\frac{d^mx_i}{dt^m}\right)^2\right>
\end{align}  Using the equations of motion $$\frac{dx_i}{dt} = v_i(\vec
x)$$ the terms $(d/dt)^m x_i$ can be related to functions on phase
space.  For example $$\frac{d^2x_i}{dt^2}= \frac{dv_i}{dt}
 = \sum_j v_j(\vec x)\frac{\partial v_i(\vec x)}{\partial x_j}$$ so that $$\alpha_2
 = \left<\left(\frac{d^2x_i}{dt^2}\right)^2\right> = \int d\mu(\vec
 X)\left(\sum_j v_j(\vec X)\frac{\partial v_i(\vec X)}{\partial
 X_j}\right)^2$$  Since $F_\Delta(\vec X)$ can not have
 singularities on the real axis (the real solutions of the equations of
 motion are assumed to be non-singular), the series \eqref{sers}
has a finite radius of convergence.  We propose to evaluate
$G(\Delta)$ by Monte-Carlo integration to evaluate the coefficients
$\alpha_m$, followed by a Pade resummation of \eqref{sers}.

For the dynamical system \eqref{dyns},  the moments $\alpha_m$ are
evaluated with respect to the measure $d\mu(\vec x)=dx dy dz \tilde
\rho(\vec x)$
\begin{align}\label{polydens}
&\tilde\rho=1-x^4-y^2-z^6 \,\, {\rm for}\,\, x^4+y^2+z^6<1\, ,\nonumber \\
&\tilde\rho=0\,\, {\rm for}\,\, x^4+y^2+z^6 \ge 1\, ,
\end{align}
We have calculated $\alpha_0\cdots\alpha_8$ for the auto-correlation
$G_x(\Delta)$ using Monte-Carlo simulation, together with a
significant amount of algebra which was also carried out with a
computer.  The corresponding $(4|4)$ Pade approximant is a ratio of
polynomials of degree 4 whose first 9 Taylor-Mclaurin series
coefficients are $\alpha_0\cdots\alpha_8$. The $(4|4)$ Pade
approximant to $G(\Delta)$ is plotted in figure 2, along with the
9'th order Taylor Mclauren series result and the result of direct
numerical simulation by a long run of the dynamical system. Note
that the 9-th order Taylor-Mclaurin series begins to differ markedly
from the result of direct numerical simultation at $\Delta\sim
0.35$, well before the first zero, whereas the $(4|4)$ Pade
approximant gives accurate result for much larger values of
$\Delta$, and is an acceptable approximation up to a neighborhood of
the first zero, $\Delta\sim 2$. Higher order Pade approximants  are
needed to obtain more real zeros, and to provide a good
approximation of the exact result for larger values of $\Delta$.
This requires more computing power than we have presently applied to
the problem. Nevertheless, the initial results are very encouraging,
suggesting that the auto-correlation may be computed without a
direct simulation of the dynamical system.

\begin{figure}[H]\centerline{
\includegraphics[width=75mm,height=75mm]{MCAuto1.eps}}
\centerline{\includegraphics[width=75mm,height=75mm]{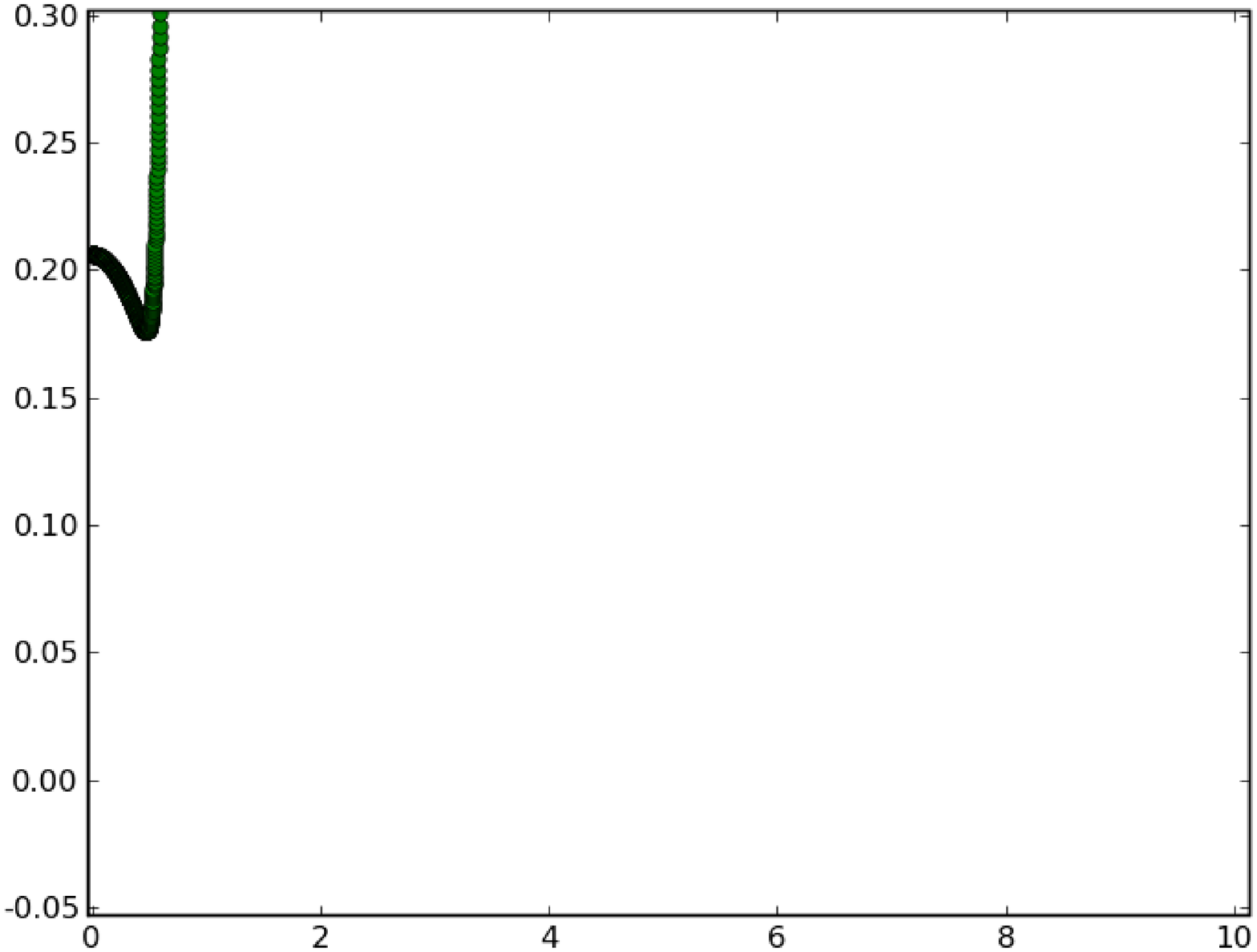}\qquad
\includegraphics[width=75mm,height=75mm]{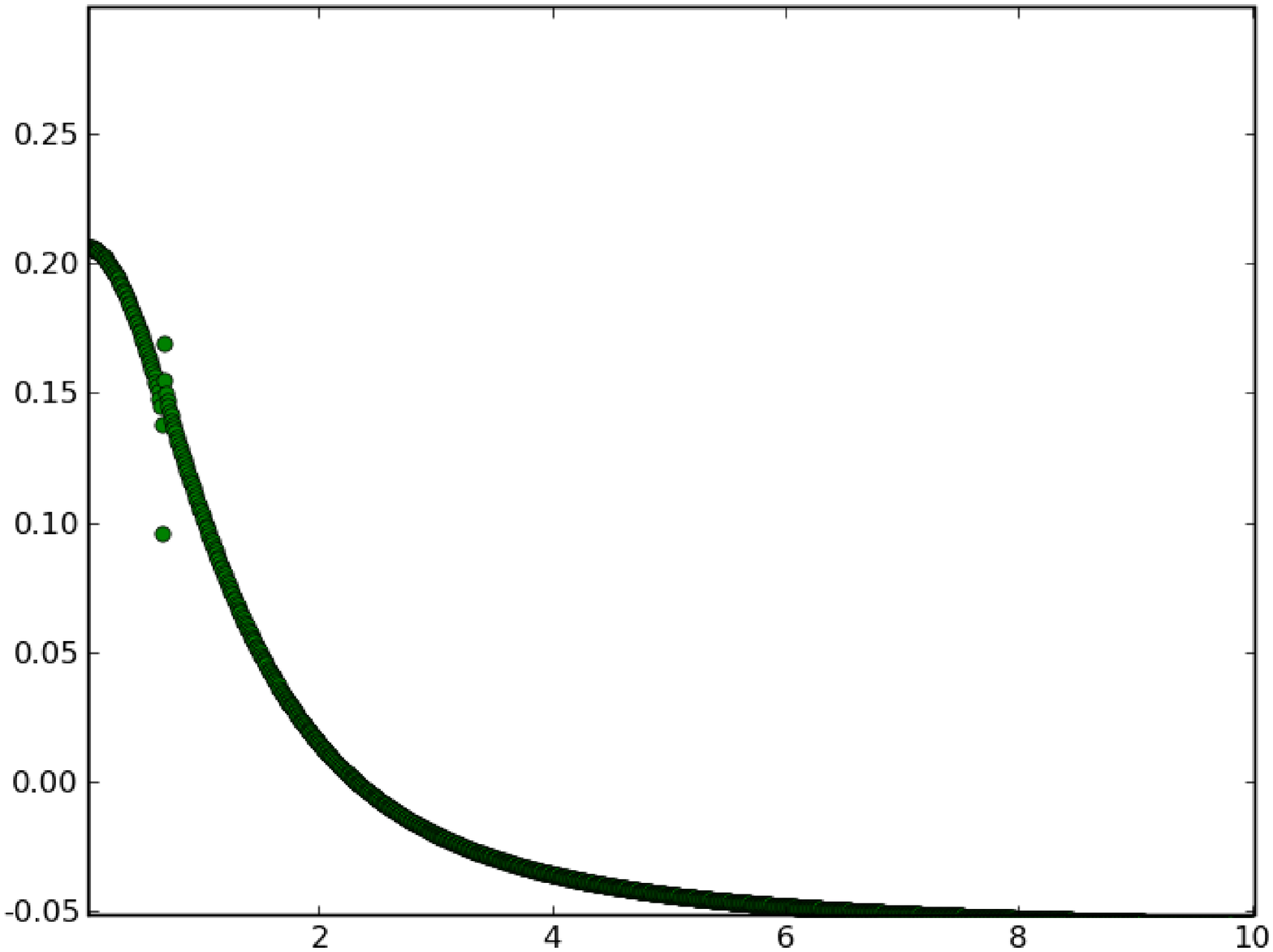}} \caption{The
auto-correlation calculated by long-run numerical simulation  and by
Monte-Carlo short run simulations (top), by 9'th order
Monte-Carlo/Taylor-Mclauren series (bottom left) ,  and by
Monte-Carlo/$(4|4)$ Pade approximant (bottom right).  The Pade
result is a smooth continuous curve, with the exception of a very
small neighborhood of the point $\Delta =0.6529$, at which there is
a spurious pole.}
\end{figure}

An oft stated property of chaotic power spectra is that there is
non-zero exponentially small component at high frequency
\cite{Frisch,Sigeti}, $\tilde G(\omega)\sim \exp(-\alpha\omega)$.
The time-scale $\alpha$ is determined by the proximity of the
nearest singularity of the auto-correlation $G(\Delta)$ to the real
$\Delta$ axis. Note that there can not be any singularities on the
real $\Delta$ axis, as it is assumed that the time evolution of the
dynamical system does not encounter any singularities.  Due to the
tendency of chaotic systems to `forget' their intial conditions, one
expects singularities of $G(\Delta)$ near the real axis to occur for
small values of $Re(\Delta)$.  This suggest that the high frequency
behavior of the power spectral density can be extracted from the
small $\Delta$ behavior of the auto-correlation.

It may be possible to use the Pade approximant to the
auto-correlation to get an estimate of the parameter $\alpha$. The
poles of the Pade approximant do not necessarily correspond to the
true analytic structure, and may in general be spurious.  In fact
the $(4|4)$ Pade approximant we have computed here has two poles
which are likely spurious, including one on the positive real axis
which is definitely spurious.  These poles are extremely close to
zeros of the Pade approximant.  There is one pole which is not near
any zero, at $\Delta=1.24 i$. Verifying that this yields a good
first approximation to $\alpha$ requires consideration of higher
orders in the Pade sequence, $(n|n)$ for $n>4$, which we have yet to
attempt.

\section{Remarks on the Fourier transform of a chaotic signal}

The Fourier transform of a chaotic trajectory $x(t)$,
\begin{align}\label{fouriert}
{\tilde x_i}(\omega) = \lim_{\epsilon\rightarrow
0}\int_{-\infty}^\infty\, dt\, e^{ i \omega t} x_i(t)\, .
\end{align}
is not well defined. Nevertheless,  it is quite common to take the
Fourier (or discrete Fourier) transform of chaotic time series with
some window of fixed duration,  squaring the amplitude to give a
kind of power spectrum.

Let us consider a function on phase space $f_\omega(\vec X)$,
defined by
\begin{align}f_\omega(\vec X) =
\frac{\omega}{2\pi}\int_0^{2\pi/\omega} dt\,x_i(t)e^{-i\omega t}
\end{align}
where \begin{align}\vec x(t=0) = \vec X\end{align} and $\vec x(t)$
satisfies the equations of motion. Next consider a chaotic
trajectory $\vec X(t)$, and define $<f_\omega>$ to be the time
averaged value of $f_\omega(\vec X(t))$ on this trajectory.  The
average can be computed by summing over values of $f_\omega(\vec
X(t))$ at arbitrary regular time intervals $\Delta t$,  yielding a
result which is independent of the interval. If one chooses the
interval $\Delta t= 2\pi/\omega$, one arrives at the formal result
\begin{align}\label{four}
<f_\omega> = \lim_{T\rightarrow\infty}\frac{1}{T}\int_0^T\, dt\,
e^{-i\omega t} x_i(t)
\end{align}
In terms of the invariant measure $d\mu(\vec X)$ on the phase space
of a chaotic system, ergodicity implies
\begin{align}\label{ergd}
<f_\omega> = \int d\mu(\vec X) f_\omega(\vec X)
\end{align}
Note that different initial conditions $\vec x(0)$ which are
different points on the same chaotic trajectory lead to a different
overall phase in \eqref{four}. On the other hand, \eqref{ergd}, too
which \eqref{four} is formally equivalent,  has no reference to an
initial condition.  This is consistent only if $<f_\omega> =0$.

It is interesting to try to compute \eqref{ergd} by Monte-Carlo
simulation. Note that any non-zero result is pure error, which we
shall see is closely related to the power spectral density.  The
Monte-Carlo calculation amounts to generating a set of $n$ initial
conditions randomly, according to the distribution of the chaotic
invariant set, and then running the dynamical system from each
initial condition for a duration $2\pi/\omega$. Given the chaotic
`loss of information' about initial conditions with time, one
expects this to yield a very similar result to direct numerical
simulation from a single initial condition over a duration $2\pi
n/\omega$.

Calculating $<f_\omega>$ by Monte-Carlo simulation, any non-zero
result is pure error,  the size of which is related to the variance
of $f_\omega$. The variance of $f_\omega$ can be expressed in terms
of the auto-correlation,
\begin{align}
<|f_\omega -<f_\omega>|^2> &= \int_0^{2\pi/\omega} dt
\int_0^{2\pi/\omega} dt' \int d\mu(\vec
X)x(t)x(t')e^{i\omega(t-t')}\\ \vec x(t=0) &= \vec X\end{align} so
that
\begin{align}\label{alm} <|f_\omega -<f_\omega>|^2>=  \int_0^{2\pi/\omega} dt
\int_0^{2\pi/\omega} dt'  G(t-t')e^{i\omega(t-t')}
\end{align}
Note that \eqref{alm} is very similar to the power spectral density,
with the exception of the boundaries of integration.

We have calculated a `power spectrum' for the dynamical system
defined by \eqref{deft},  defined as the square of the amplitude of
a discrete Fourier transform within some time window,  or of a
Monte-Carlo approximation to $f_\omega$.  The results of both
calculations for $\beta=1$ are plotted in figure 3.

\begin{figure}[H]\centerline{
\includegraphics[width=60mm,height=60mm]{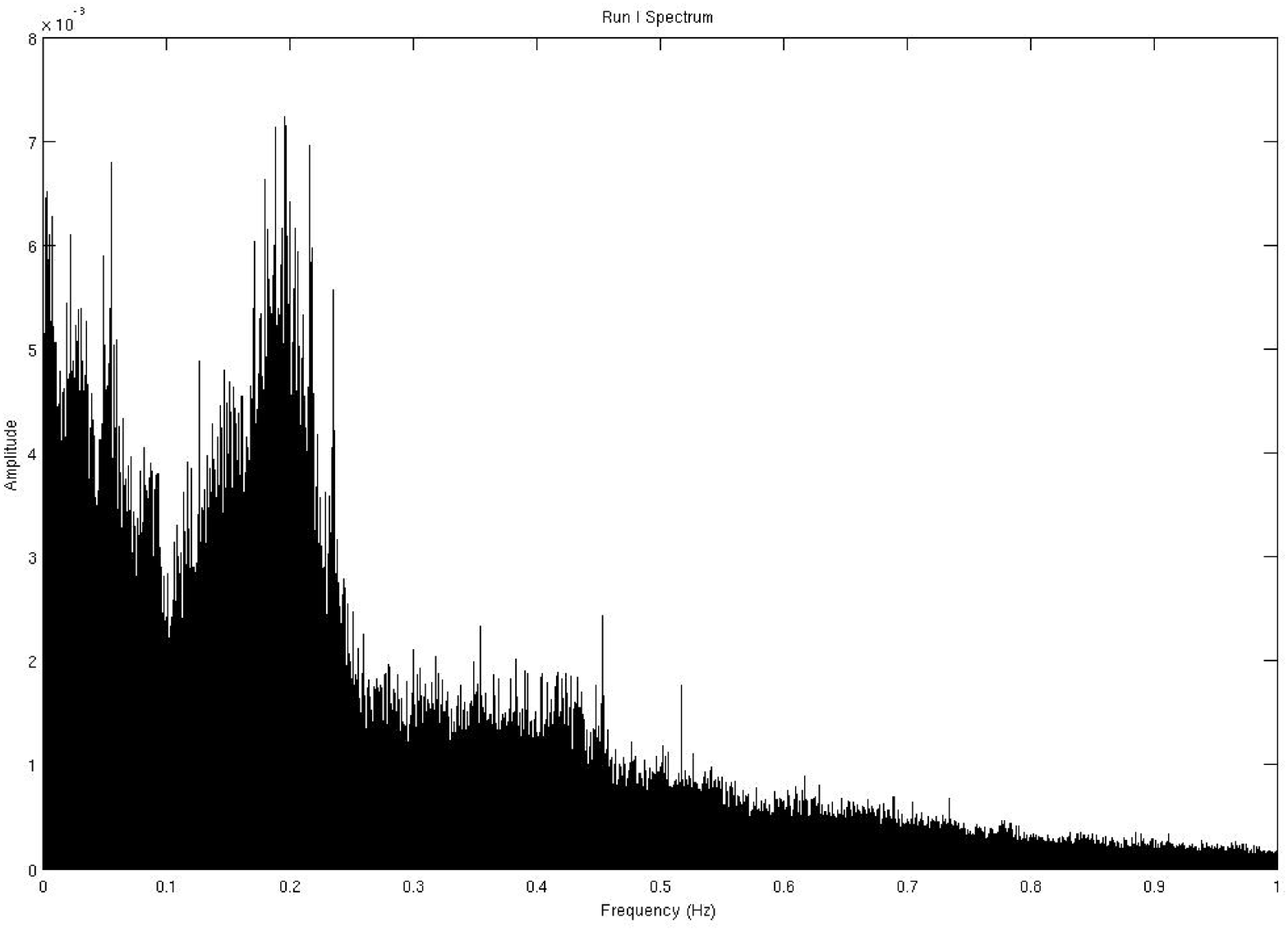}\qquad
\includegraphics[width=60mm,height=60mm, angle=90]{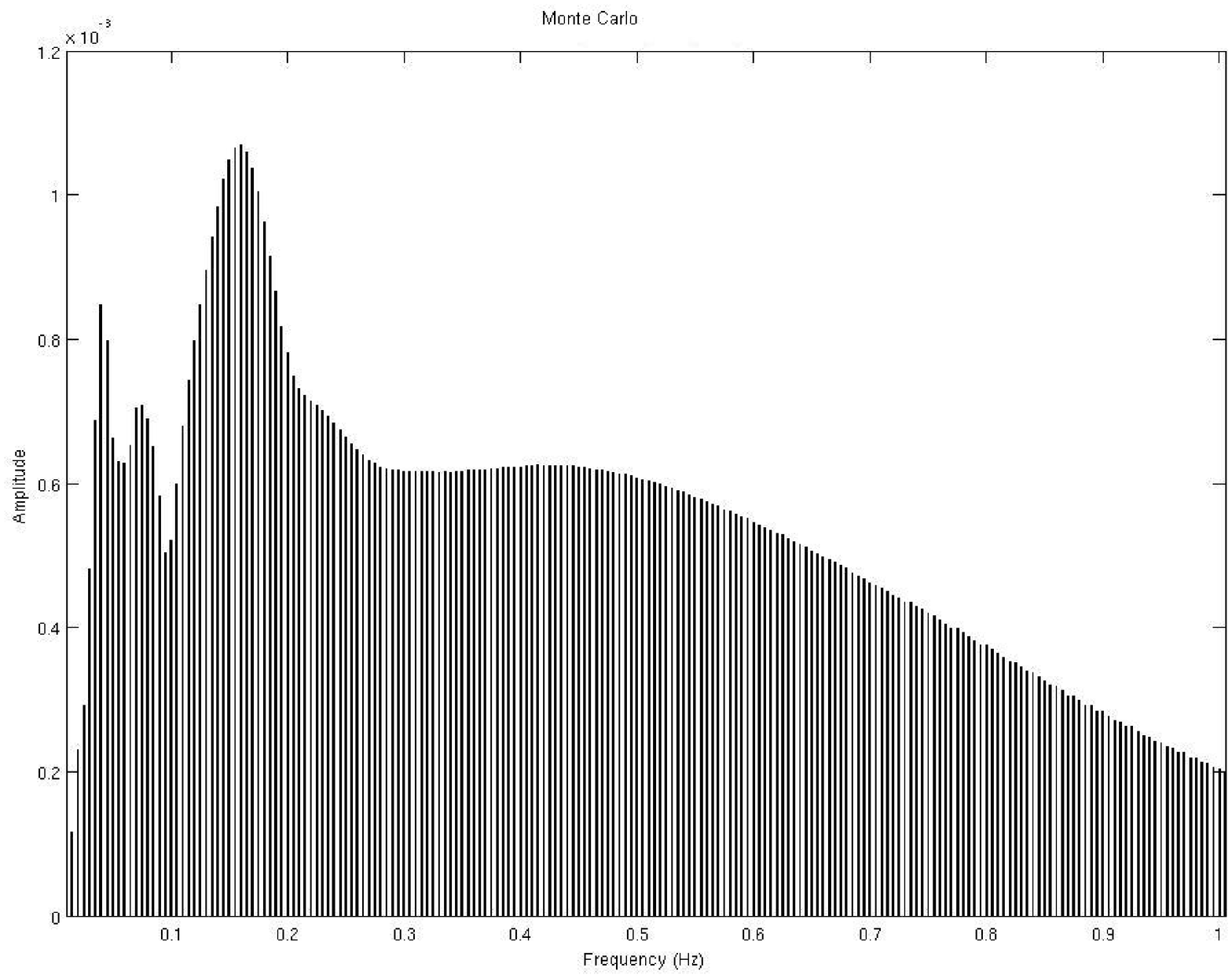}}
\caption{`Power spectrum' computed by Fast Fourier Transform of a
chaotic signal (on the left) and by Monte-Carlo simulation (on the
right). }
\end{figure}

Due to the tendency of chaotic systems to `forget' their initial
conditions, on might expect the Fourier transform of a run of finite
but long duration to be qualitatively similar to the Monte-Carlo
result,  which involves shorter runs from randomly generated initial
conditions.  Indeed the results plotted in figure 3 have some crude
qualitative equivalence.   In some sense, both these results can
be viewed as nothing but finite total simulation time errors, the
exact result vanishing identically, where the error is related to a
quantity \eqref{alm} similar, but not equivalent, to the power
spectral density.
%Note for example that the `power spectrum' in figure 4 (for
%$\beta=1.5$) has a pronounced peak at 3 Hz (OR IS IT 0.3??? THIS
%LOOKS LIKE 0.3 BUT FIGURE 1 LOOKS LIKE 3.0 ...???), which can be
%clearly read from the (better defined) auto-correlation in figure 1.

\section{Conclusions}

Direct numerical simulation of chaotic dynamical systems is a viable
method to compute their statistics.  However this approach rarely
yields theoretical insight, and places extreme or prohibitive
demands on computational resources for systems with a very large
number of degrees of freedom.  The intent of this work has been to
develop an inverse method, introduced in \cite{Zach}, which permits
the calculation of statistical quantities of chaotic systems without
the use of direct numerical simulation over long time durations.

The inverse method generates chaotic dynamical systems, given an
invariant measure and a two-form.  At present we we have tested the
approach on systems with a small number of degrees of freedom, by
computing static statistical quantities in \cite{Zach} such as equal
time correlations, and time dependent statistical quantities such as
auto-correlation functions in the present article.

Already at the level of a few degrees of freedom, one should be able
to make theoretical progress which is not possible with studies
involving only direct numerical simulation.  In particular it should
be possible to construct a classification of chaotic dynamical
systems according to the phase space analytic structure of an
invariant distribution and a two-form.

Given the invariant measure of a chaotic dynamical system, which is
the starting point of the inverse approach,  one can write closed
form expressions for equal time correlation functions, for which
analytic approximation schemes may exist, such as saddle point
expansions.  Non-equal time correlation functions, such as
auto-correlations,  may also be computable without recourse to
direct numerical simulation, by Pade re-summation of short time
expansions. In fact,  many of the calculations arising using the
inverse approach bear a close resemblance to calculations in quantum
field theory,  which may yield useful ideas in the present context
of chaotic dynamics.

The true power of the inverse method will come for a large number of
degrees of freedom,  since this approach is amenable to parallel
computation. There are numerous interesting problems which remain to
be solved,  in particular that of reverse engineering a chaotic
dynamical system of particular physical interest, such as a
turbulent fluid.  Efforts along these lines are underway.

There remain subtle and interesting theoretical problems concerning
the meaning of the invariant distribution, which is the starting
point of the inverse method, when the information dimension is
fractional. These subtleties have been discussed in \cite{Zach} and
remain to be resolved.

\section{Acknowledgments}

We wish to think S. Libby for helpful discussions.  The work of G.
Guralnik and C. Pehlevan is supported in part by funds provided by
the U.S. Department of Energy (DoE) under DE-FG02- 91ER40688-Task D.

\end{document}